\documentclass{iitpressproc}

\usepackage{graphicx}
\usepackage{amsmath,amssymb,amsfonts}

\begin{document}

\title{Recent results from the search for the critical point of strongly
interacting matter at the CERN SPS}

\author{Peter Seyboth
\address{Max-Planck Institut f\"ur Physik, Munich, Germany\\
and\\ 
Jan Kochanowski University, Kielce, Poland}\\[2ex]
for the NA49 and NA61/SHINE collaborations
}

\maketitle

\begin{abstract}
Recent searches at the CERN SPS for evidence of the critical point
of strongly interacting matter are discussed. Experimental results on theroretically 
expected signatures, such as event-to-event fluctuations of the particle multiplicity
and the average transverse momentum as well as intermittency in particle production
are presented. 
\end{abstract}


\section{Introduction}
\label{Introduction}

Exploration of the phases of strongly interacting matter is the main
purpose of the study of high energy heavy-ion collisions. Theoretical
considerations~\cite{qgp_review} suggest that the phase boundary between confined hadrons
at low and quasi-free quarks and gluons at high temperature and/or density 
is of the first order in systems with large net-baryon density (or equivalently
baryochemical potential $\mu_B \gg 0$). 
Lattice QCD calculations~\cite{lattice_review} can provide quantitative predictions at zero
net baryon density ($\mu_B = 0$) and find that the transition is a rapid crossover. 
Thus a critical point is expected as the endpoint of the first-order
transition line. However, lattice QCD is not yet able to cope with $\mu_B > 0$
in a strict way. Predictions of the existence and location of the critical
point (CP) in the phase diagram of T versus $\mu_B$ have to be obtained from
extrapolations which arrive at conflicting results. Some find a CP in a region
accessible to experiments at the SPS and the RHIC beam energy scan~\cite{cp_acc},
others locate the CP at high $\mu_B$ where heavy-ion experiments are not able to produce the
deconfined phase~\cite{cp_nonacc} or they find no CP at all~\cite{cp_nonex}. Clearly
it is important to address this issue by experimental studies.

\begin{figure}
\centering
\includegraphics[width=0.45\columnwidth]{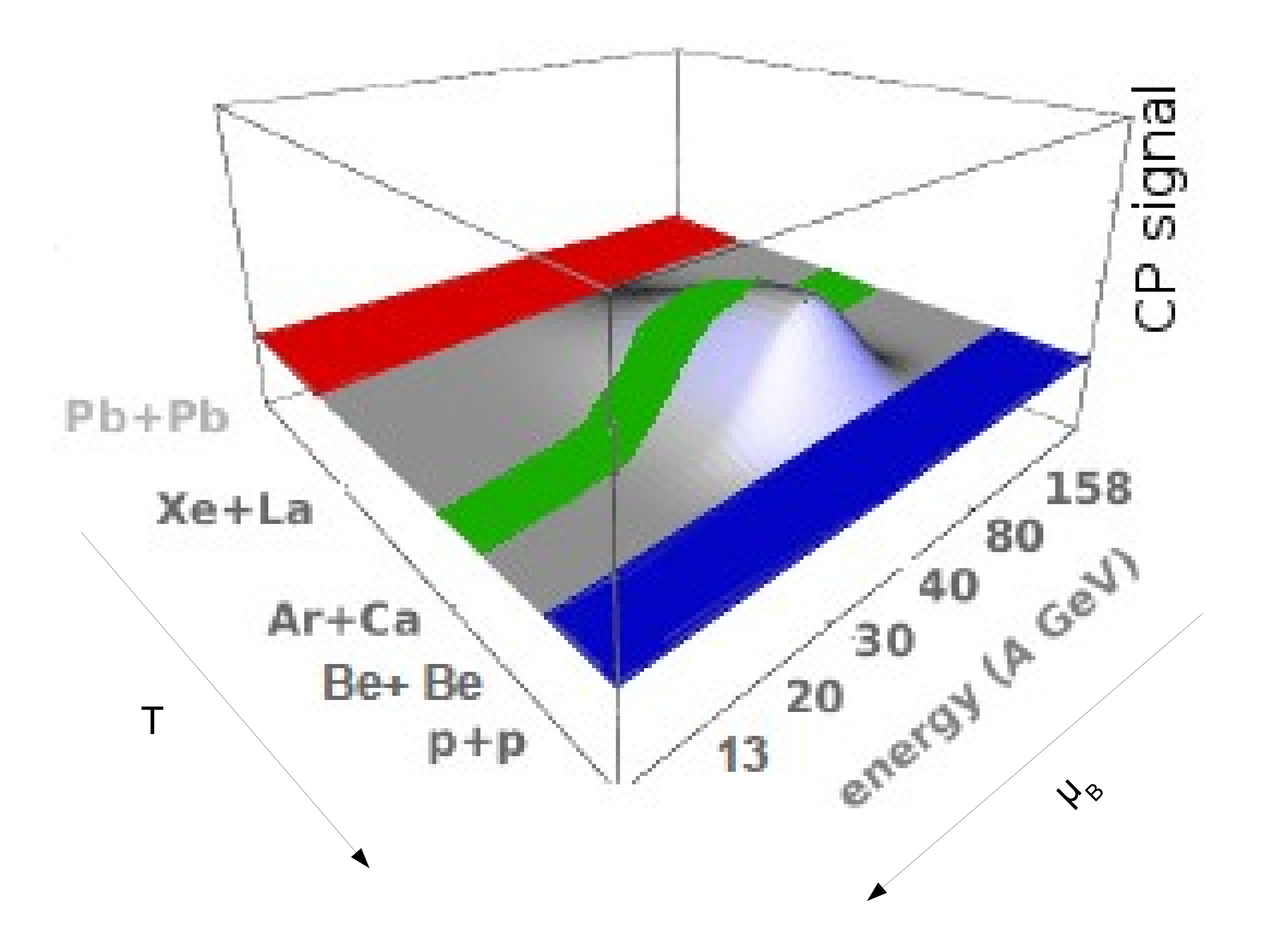}
\includegraphics[width=0.45\columnwidth]{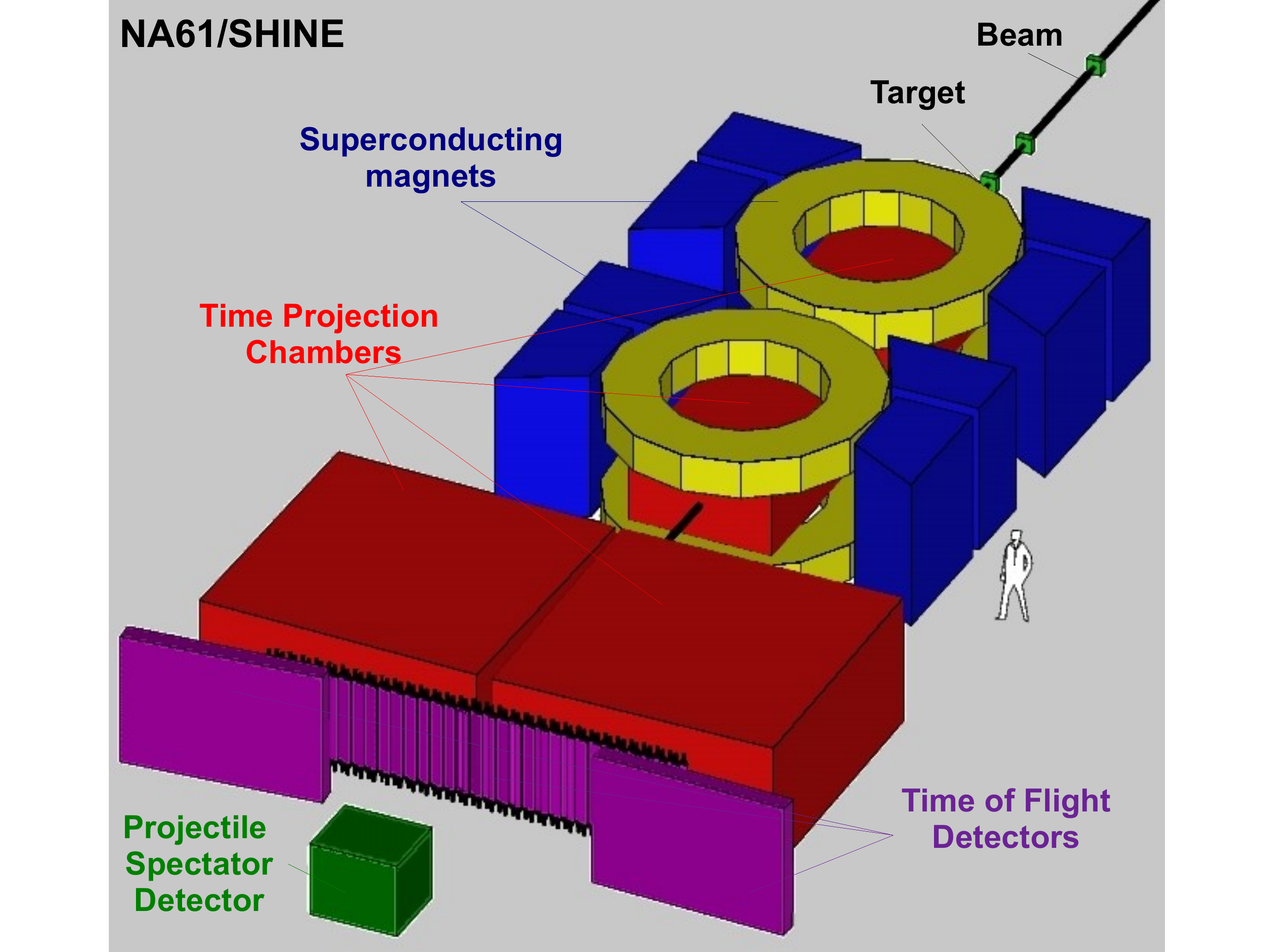}
\caption{Left: expected hill of fluctuations in a scan of the phase diagram with a critical point.
Right: schematic view of the NA49/NA61 detector}
\label{hill_det}
\end{figure}

At a CP the correlation length $\xi$ diverges leading to a strong increase of
suitable correlation measures such as event-to-event fluctuations 
of the multiplicity and average transverse momentum of produced particles~\cite{cp_signals} as well as
local density fluctuations resulting in the appearance of intermittency in particle 
production~\cite{antoniou_int}. Owing to the finite size and short lifetime of the fireballs
produced in collisions of nuclei, $\xi$ is expected not
to exceed 3-6 fm in Pb+Pb collisions. Moreover, correlations may be diluted by rescattering
of the produced particles before final freezeout.  

A scan of the phase diagram by varying the sizes of colliding nuclei (change of
rescattering probabilty) and energies of the collisions (change of $\mu_B$) is a promising
search strategy~(see Fig.~\ref{hill_det}, left). A coinciding maximum of several fluctuation measures 
would indicate the existence and the location of the CP. This program was started by the 
NA49 collaboration~\cite{na49} and is now pursued systematically by the NA61/SHINE experiment~\cite{na61}.

\section{Detector and recorded data}
\label{detector}

The NA61 experiment, the successor of NA49, uses a fixed target spectrometer with particle
identification, covering mainly the forward region in the center-of-mass rapidity.
The schemtic view in Fig.~\ref{hill_det}~(right) shows the system of four large Time Projection
Chambers for particle tracking and momentum measurement. The first two 
are placed inside superconducting magnets with combined bending power
of 9 Tm. Particle identification is obtained by measuring the energy loss by ionisation 
in the gas of the TPCs with precision of about 4 \% and the time-of-flight
in scintillation counter walls with resolution of 60-80 ps. For the NA61 program the NA49 detector
was upgraded by a new cellular zero degree calorimeter (Projectile Spectator Detector)
with single beam nucleon energy resolution and a He filled beam pipe trough the TPCs
to reduce beam induced $\delta$-ray background. Finally, the digital part of the TPC readout
was replaced resulting in a factor 10 increase of the data acquisition rate.

NA49 recorded data on central Pb+Pb collisions at a set of energies (20$A$, 30$A$, 40$A$, 80$A$ and 158$A$ GeV)
through the SPS energy range in the period 1994 - 2002. Additionally a smaller set of data was
taken for C+C and Si+Si collisions at 40$A$ and 158$A$~GeV. NA61 expands this program and
resumed in 2009 a comrehensive scan of energies (13$A$ GeV + NA49 energies) 
and nuclear sizes (p+p, Be+Be, Ar+Ca, Xe+La, Pb+Pb) which has been completed
for the lightest two systems. 

\begin{figure}
\centering
\includegraphics[width=0.80\columnwidth]{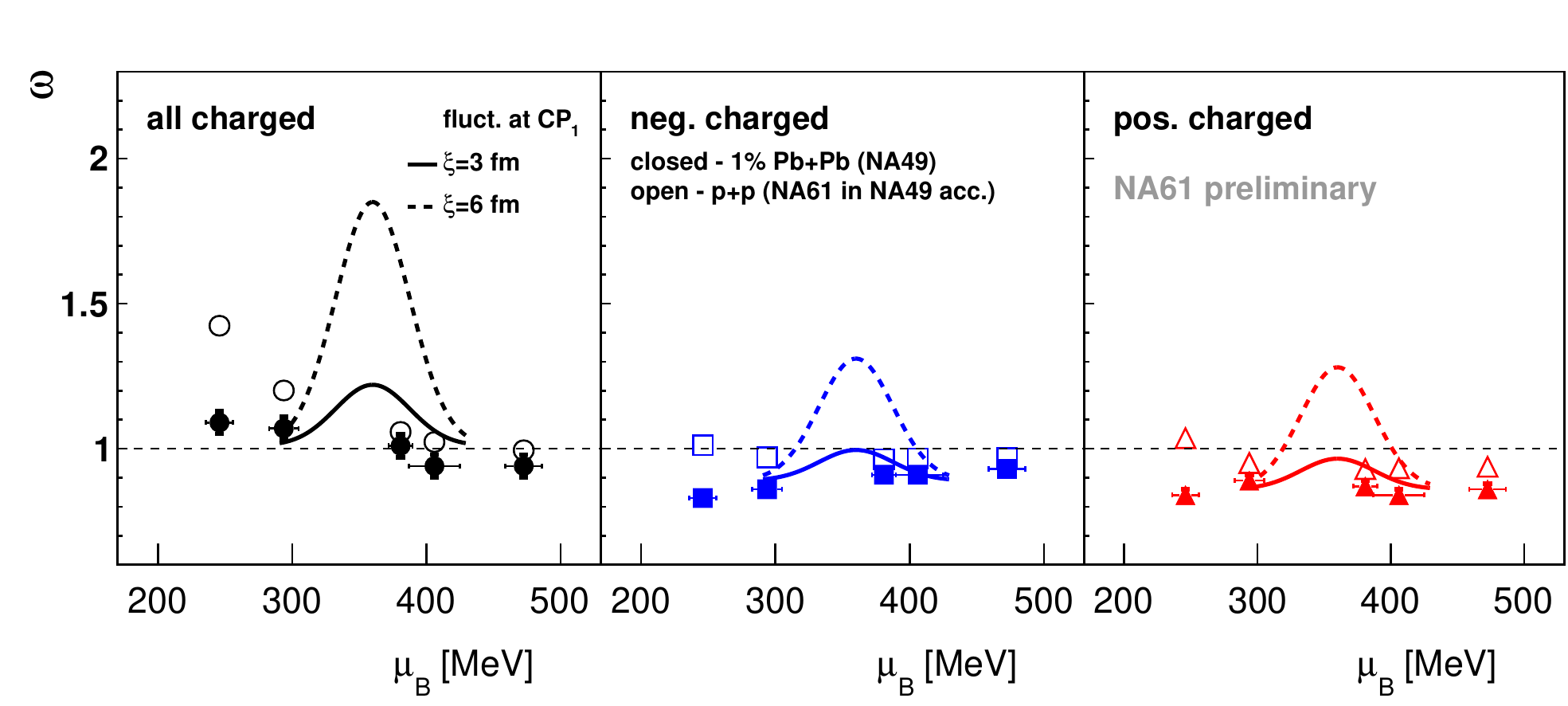}
\includegraphics[width=0.80\columnwidth]{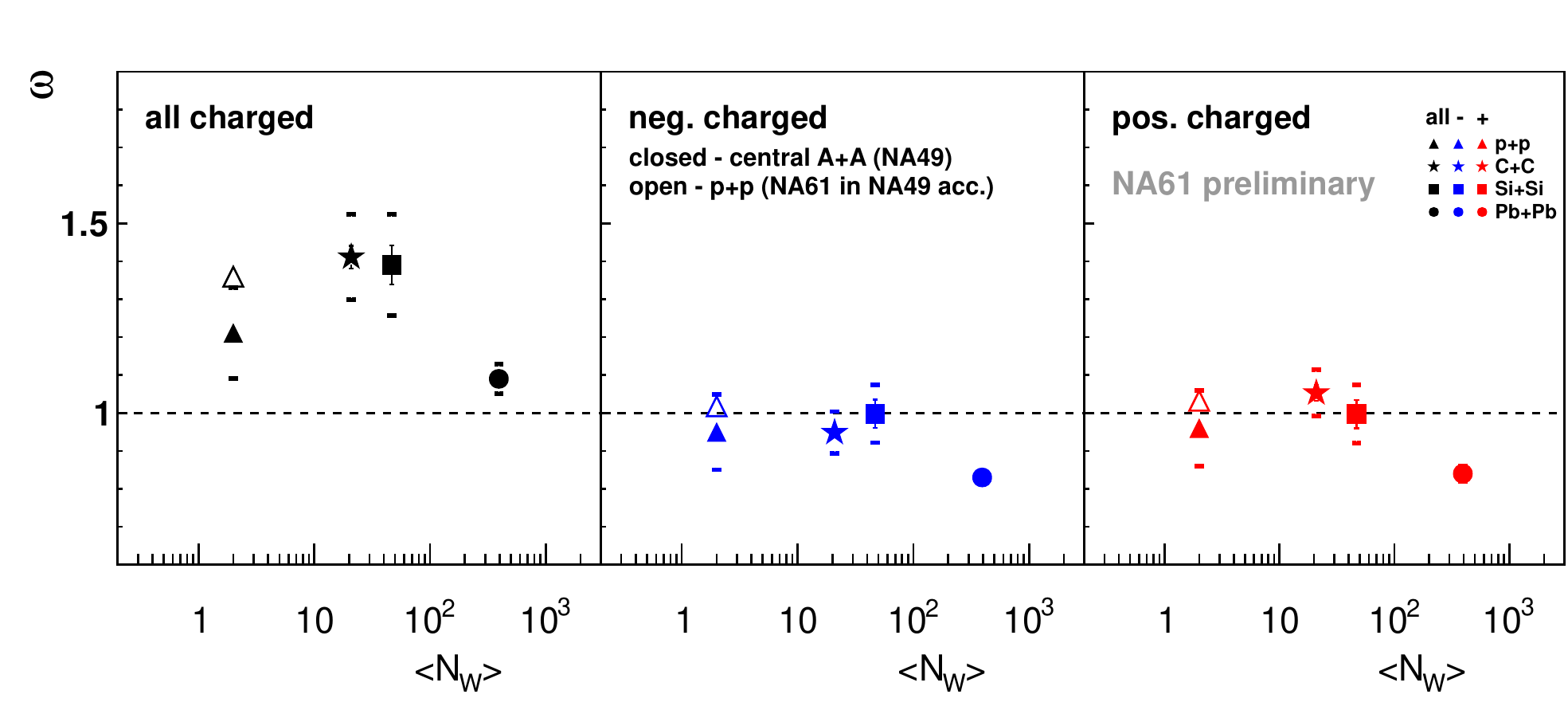}
\caption{Scaled variance $\omega$ of the multiplicity distribution of charged particles.
Top: versus $\mu_B$ for the 1~\% most central Pb+Pb collisions and inelastic p+p reactions
for $1.0 < y < y_{beam}$ (assuming the pion mass). Bottom: versus the number of wounded nucleons $N_W$
in inelastic p+p ($1.1 < y < 2.6$) and the 1~\% most central C+C, Si+Si and Pb+Pb collisions 
at 158$A$ GeV ($1.0 < y < y_{beam}$). Full symbols show results of NA49~\cite{omega_mub}, open symbols 
NA61 (preliminary).}
\label{omega}
\end{figure}

\begin{figure}
\centering
\includegraphics[width=0.80\columnwidth]{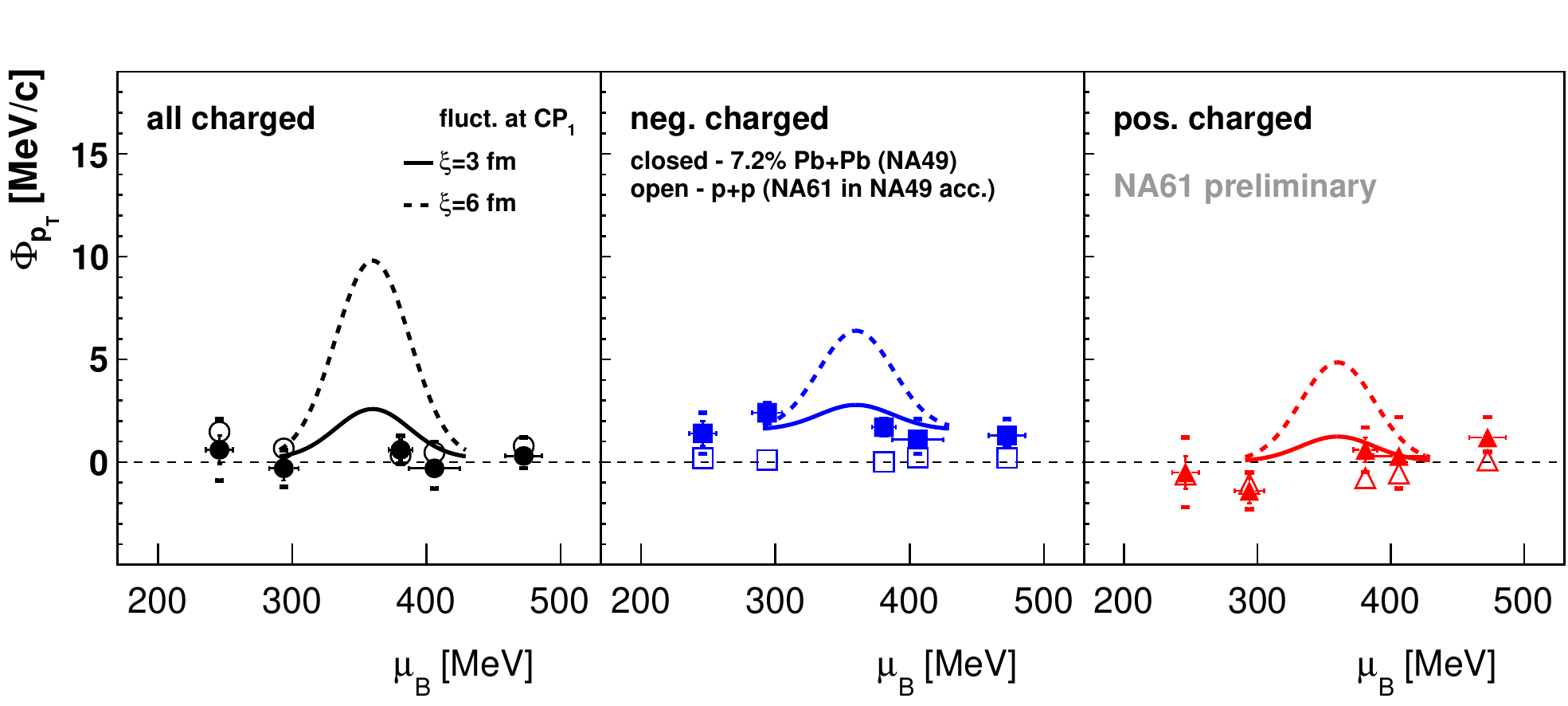}
\includegraphics[width=0.80\columnwidth]{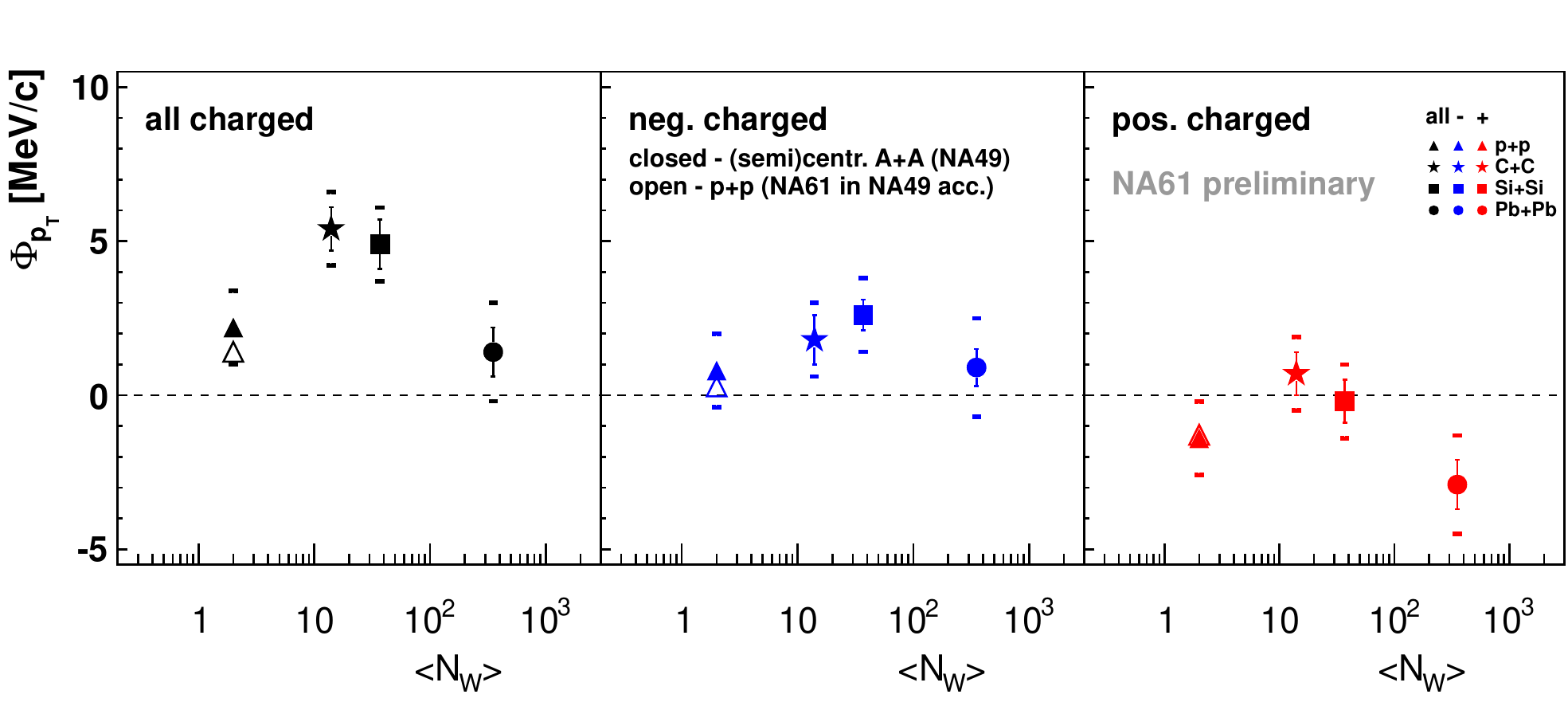}
\caption{Fluctuation measure $\Phi_{p_T}$ of the average transverse momentum of charged particles.
Top: versus $\mu_B$ for the 7.2~\% most central Pb+Pb collisions (full symbols, NA49~\cite{phipt_mub})
and inelastic p+p reactions (open symbols, NA61 preliminary). Bottom: versus the number of wounded nucleons $N_W$
in central C+C, Si+Si and Pb+Pb collisions at 158$A$ GeV (NA49~\cite{phipt_nw}) and inelastic p+p reactions
(NA61 preliminary). Results are for cms rapidity $1.1 < y < 2.6$ assuming the pion mass.}
\label{phipt}
\end{figure}

\section{Fluctuations of the particle multiplicity}
\label{mfluct}

The signature of a CP is expected to be primarily an increase of multiplicity fluctuations~\cite{cp_signals}
which are usually quantified by the scaled variance $\omega = (\langle N^2 \rangle - \langle N \rangle^2)/\langle N \rangle$
of the distribution of particle multiplicities $N$ produced in the collisions. The measure $\omega$
is "intensive", i.e. it is independent of the number of wounded nucleons $N_W$ (size or volume) of
the system in models which assume nucleus+nucleus collisions to be a superposition of
nucleon+nucleon reactions. However, $\omega$ is sensitive to the unavoidable fluctuations of $N_W$~\cite{sintensive}.
Therefore the measurements were restricted to the 1~\%  most central collisions. Results
for charged particles in Pb+Pb collisions (NA49~\cite{omega_mub}) are shown in Fig.~\ref{omega}~(top) versus $\mu_B$ (obtained
from statistical model fits to yields of different particle types at the various collision energies)
and compared to preliminary NA61 results from p+p reactions. The data do not support a maximum
as might be expected for a CP (see curves~\cite{cp_omega}). NA49 also obtained results for different 
size nuclei at the top SPS energy of 158$A$ GeV (see Fig.~\ref{omega}~(bottom)).
Here there may be an indication of a maximum for medium size nuclei. A new identification procedure
(identity method~\cite{identity}) allowed to measure the energy dependence of fluctuations of identified 
proton, kaon and pion multiplicities in p+p and Pb+Pb collisions. As in the case of charged particle multiplicities
no indication of a CP is found. It was pointed out that higher moments of the multiplicity distributions
are more sensitive to effects of the CP~\cite{high_mom}. Unfortunately the systematic uncertainties
of the measurements in NA49 and NA61 at present do not allow meaningful conclusions.

\begin{figure}
\centering
\includegraphics[width=0.80\columnwidth]{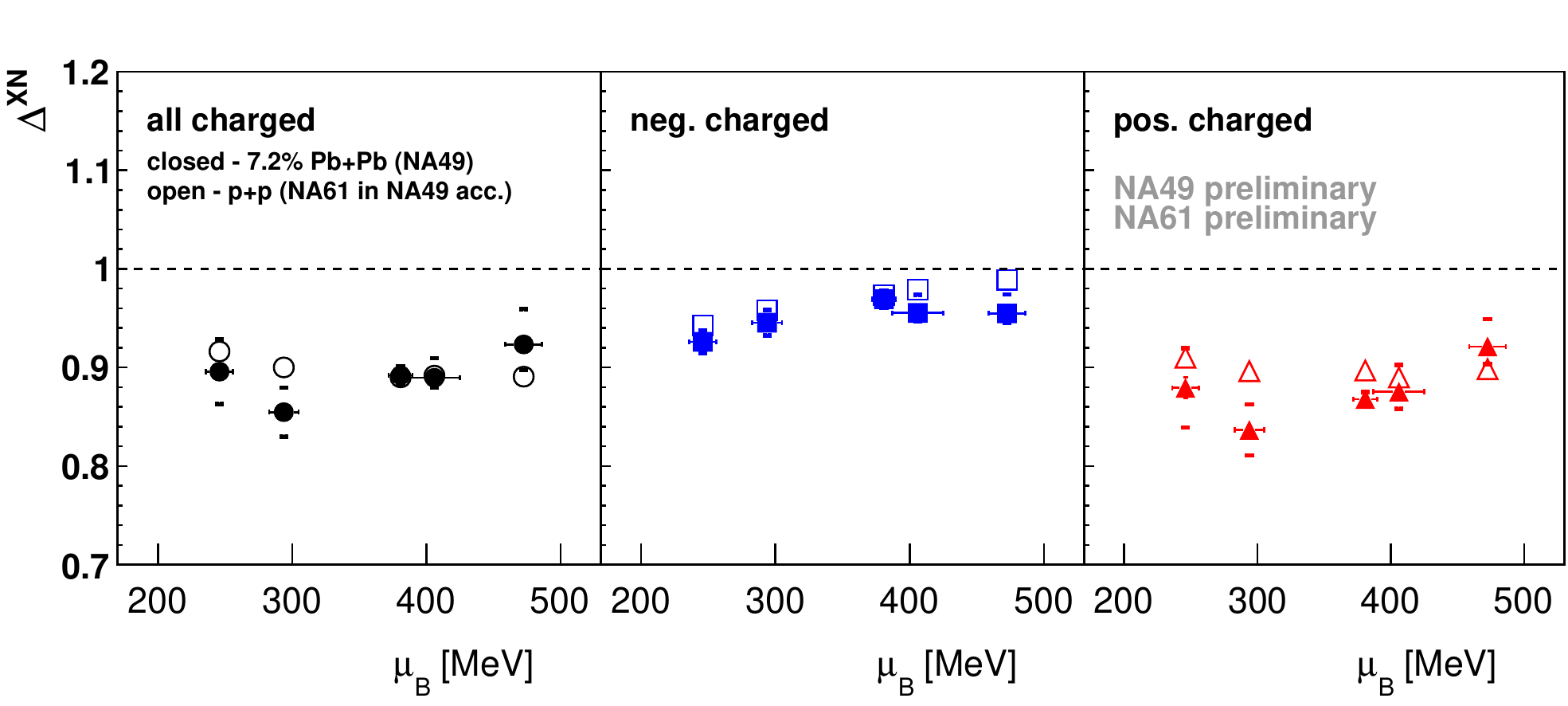}
\includegraphics[width=0.80\columnwidth]{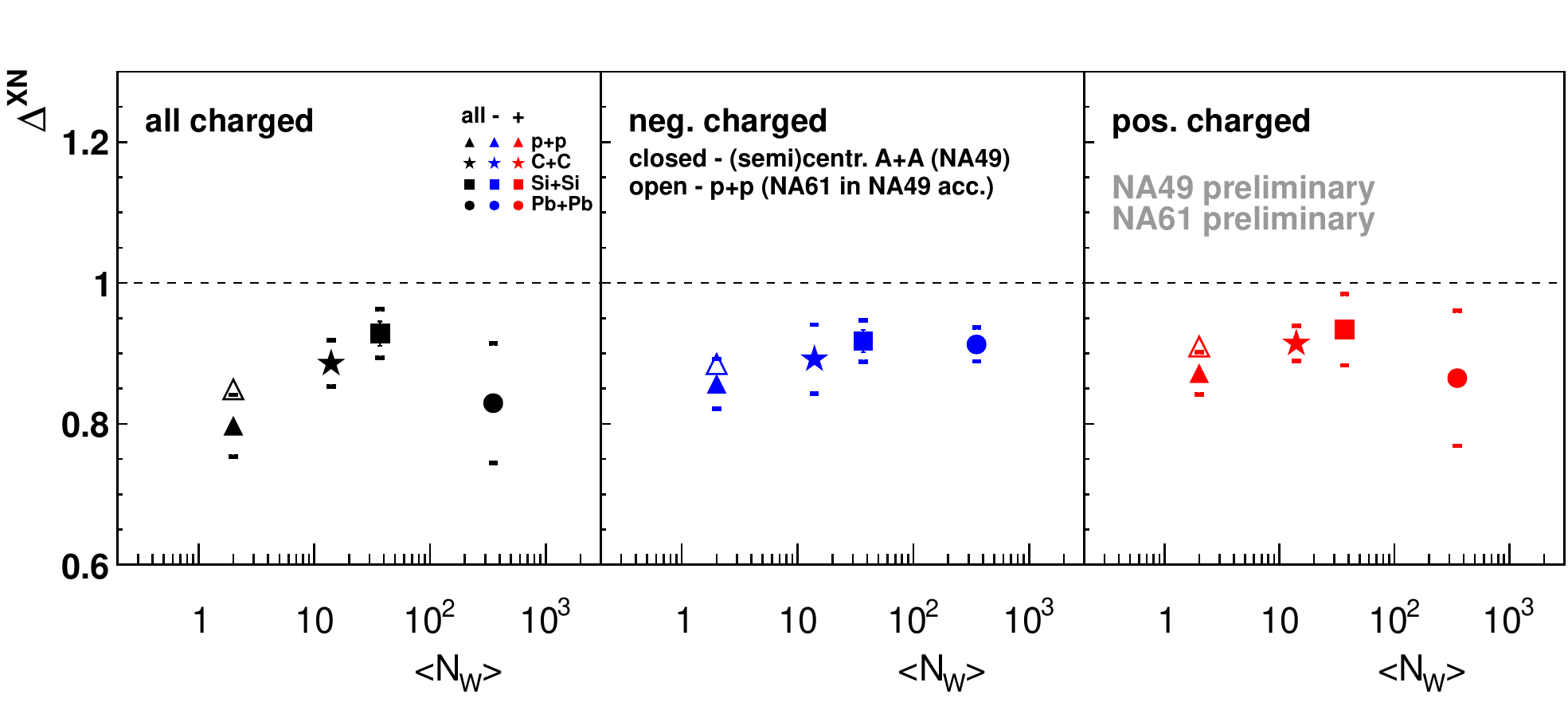}
\caption{Fluctuation measure $\Delta^{P_T,N}$ of the average transverse momentum of charged particles.
Top: versus $\mu_B$ for the 7.2~\% most central Pb+Pb collisions (full symbols)
and inelastic p+p reactions (open symbols). Bottom: versus the number of wounded nucleons $N_W$
in inelastic p+p and central C+C, Si+Si and Pb+Pb collisions at 158$A$ GeV.
Results are for cms rapidity $1.1 < y < 2.6$ assuming the pion mass. (NA49 and NA61 preliminary).}
\label{Delta}
\end{figure}

\section{Fluctuations of the average transverse momentum}
\label{ptfluct}

Enhanced fluctuations are also expected for the average transverse momentum $p_T$ when the
freezeout occurs close to the CP~\cite{cp_signals}. A suitable measure $\Phi_{p_T}$ was proposed in 
\cite{phipt}, which is "strongly intensive", i.e. independent of both $N_W$ and its fluctuations.
Results on the dependence of $\Phi_{p_T}$ on $\mu_B$ in central Pb+Pb (NA49~\cite{phipt_mub}) and
inelastic p+p collisions (NA61 preliminary) are plotted in Fig.~\ref{phipt}~(top) and compared
to expectations for a CP~(curves in Fig.~\ref{phipt}~(top)~\cite{cp_omega}). Measurements for different
size nuclei at the top SPS energy of 158$A$ GeV are shown in Fig.~\ref{phipt}~(bottom). As found 
for $\omega$ there is no evidence for a CP from the dependence of $\Phi_{p_T}$ on $\mu_B$, but
there may be a maximum for medium-size nuclei.

Recently a new class of strongly intensive measures was proposed in Ref.~\cite{sintensive}. Whereas
$\Sigma^{P_T,N}$ is closely related to $\Phi_{p_T}$ the quantity $\Delta^{P_T,N}$ is sensitive to
fluctuations of $p_T$ and $N$ in a different combination. Results shown in Fig.~\ref{Delta} are
inconclusive, in particular, as at present there are no predictions for the effect of a CP in this
observable.

\begin{figure}
\centering
\includegraphics[width=0.45\columnwidth]{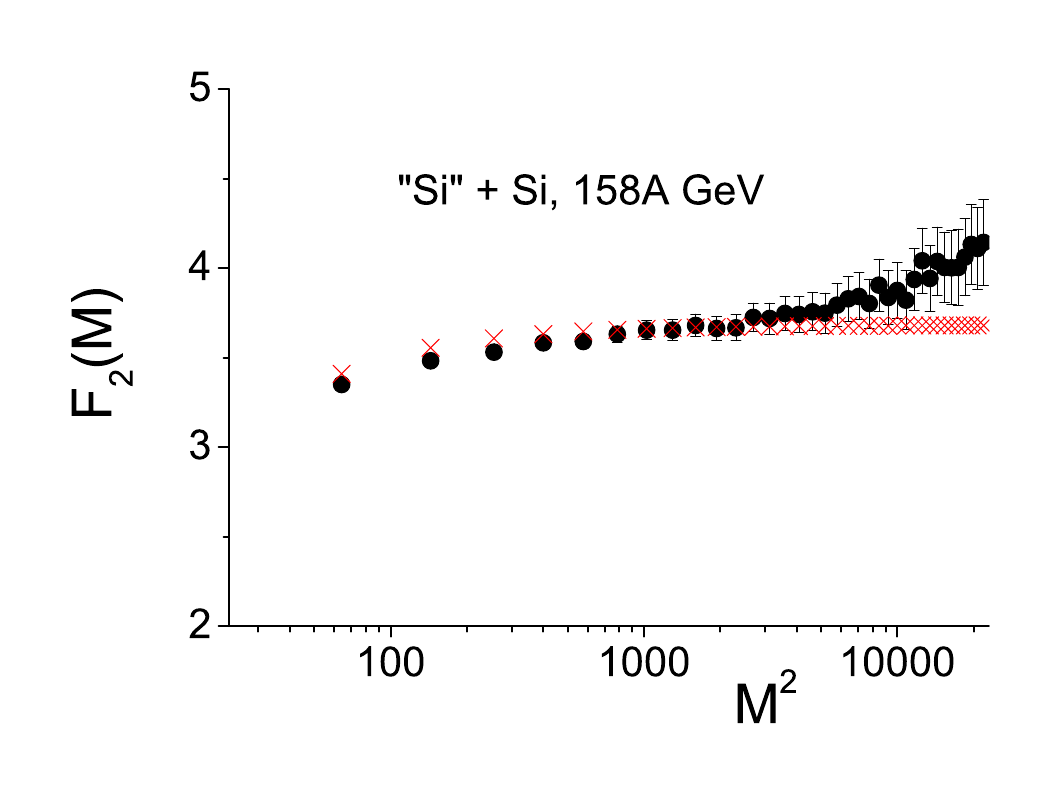}
\includegraphics[width=0.45\columnwidth]{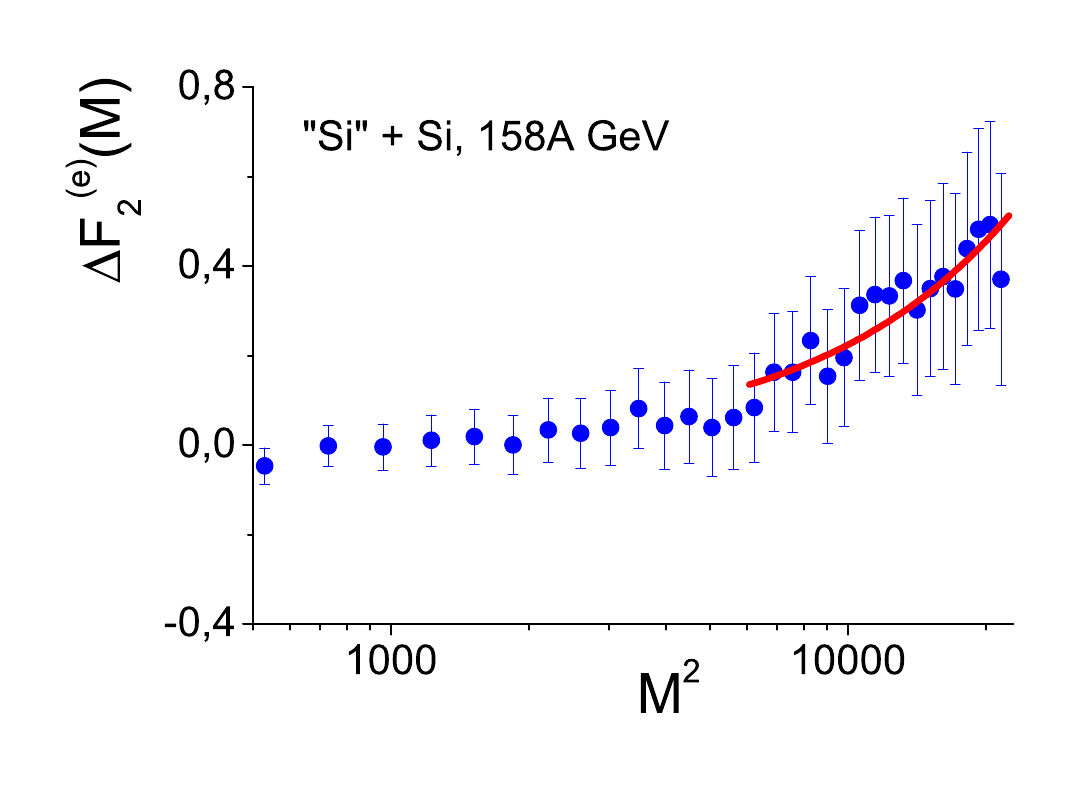}
\caption{Scaled factorial moments of protons in rapidity $|y| < 0.75$ for the 12.5~\% most central Si+Si collisions
at 158$A$ GeV (NA49~\cite{na49_pint}). Left: $F_2(M)$ versus the number of cells $M^2$ 
in transverse momentum space.
Dots show data, crosses the mixed event background. Right: $\Delta F_2(M)$ versus $M^2$; 
dots show background subtracted data, the curve the result of a power-law fit $\Delta F_2(M) \propto M^{2\Phi_2}$.}
\label{prot_int}
\end{figure}

\begin{figure}
\centering
\includegraphics[width=0.45\columnwidth]{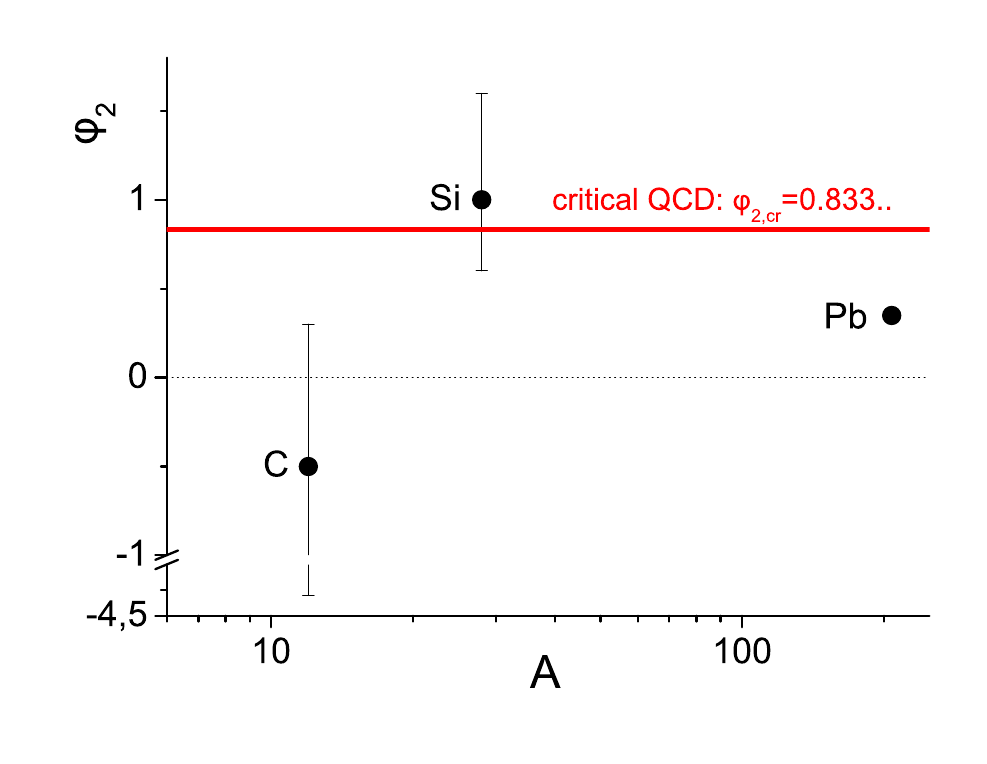}
\includegraphics[width=0.45\columnwidth]{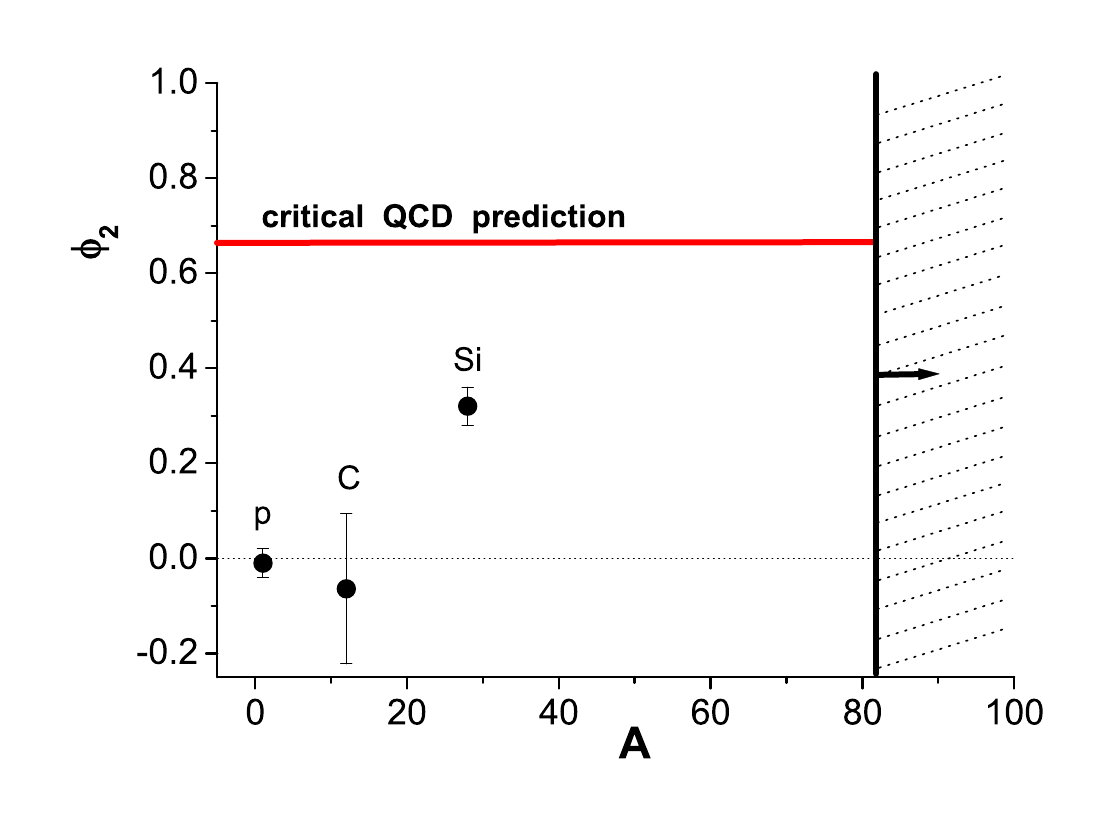}
\caption{Exponent $\Phi_2$ obtained from power law fits to second scaled factorial moments of 
protons~\cite{steph_pfluct} (left) and low-mass $\pi^+\pi^-$ pairs~\cite{na49_pipi_int} (right)
for several collision systems at 158$A$ GeV. }
\label{phi_2}
\end{figure}

\section{Local density fluctuations of protons and low-mass $\pi^+\pi^-$ pairs}
\label{factmom}

Theoretical investigations~\cite{antoniou_int} predict near the CP the appearance of local density fluctuation
for protons~\cite{steph_pfluct} and low-mass $\pi^+\pi^-$ pairs of power-law nature with 
known critical exponents~\cite{antoniou_prot}. These can
be studied by the intermittency analysis method in transverse momentum space using second factorial
moments $F_2(M)$, where $M$ is the number of subdivisions in each $p_T$ direction. After combinatorial 
background subtraction the exponents $\Phi_2$ are obtained from a 
power-law fit to the corrected moments $\Delta F_2(M) \propto M^{2\Phi_2}$. The procedure is illustrated
for protons in central Si+Si collisions in Fig.~\ref{prot_int}. The resulting values of $\Phi_2$ 
obtained for central C+C, Si+Si and Pb+Pb collisions at 158$A$~GeV~\cite{na49_pint}
are plotted in Fig.~\ref{phi_2} (left).
Remarkably, $\Phi_2$ seems to reach a maximum for Si+Si collisions which is consistent with the
theoretical expectation for the CP. A similar conclusion was reached for 
low-mass $\pi^+\pi^-$~pairs~\cite{na49_pipi_int} (see Fig.~\ref{phi_2} (right)).

\begin{figure}
\centering
\includegraphics[width=0.8\columnwidth]{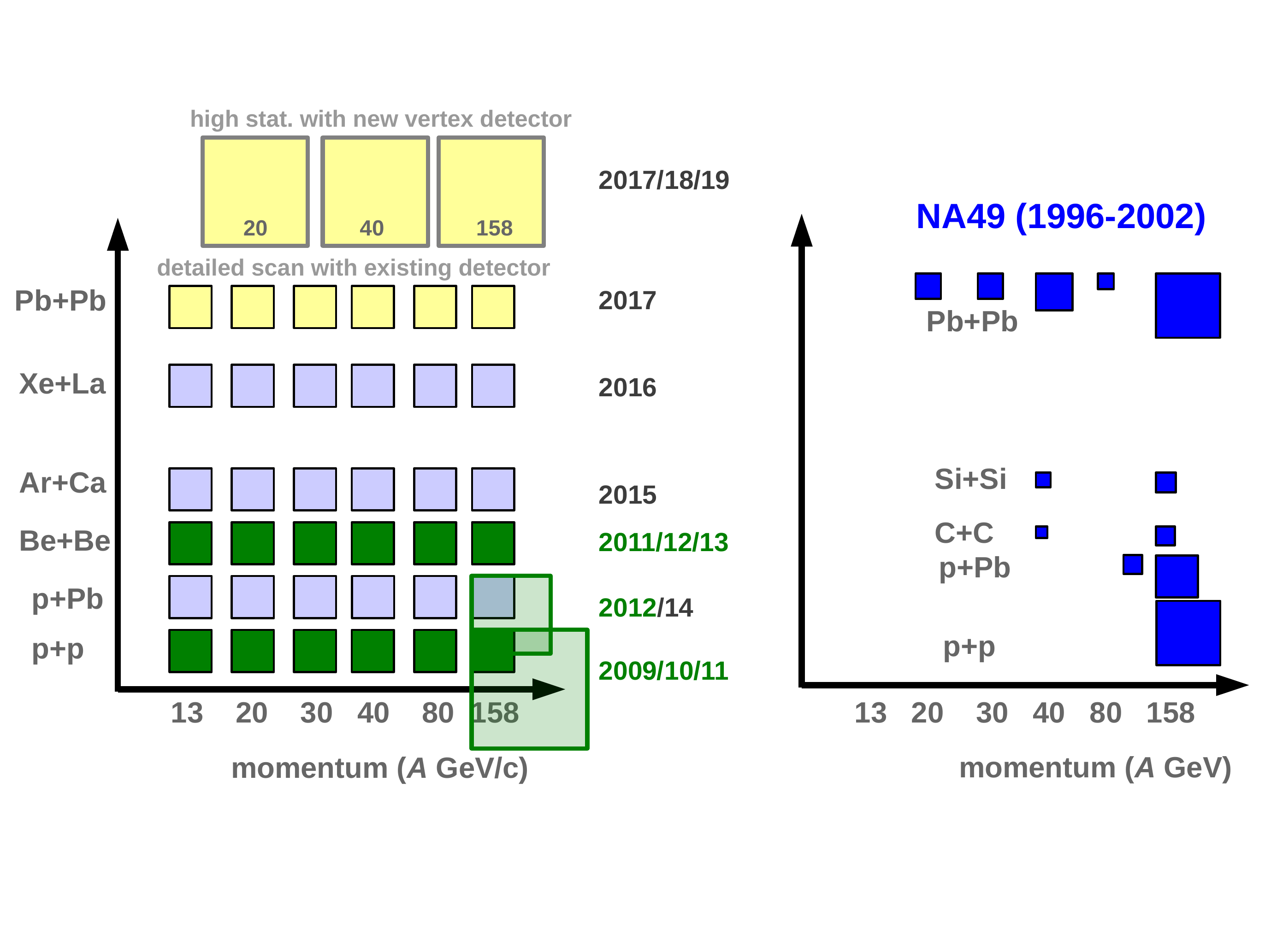}
\caption{Reactions and energies of the scan of the phase diagram by NA61 (left, in progress) 
and systems previously studied by NA49 (right). }
\label{na61_na49_scan}
\end{figure}

\section{Conclusion}

The continuing search in nucleus+nucleus collisions for the maximum of fluctuations predicted 
for a critical point of strongly interacting matter has not yet turned up firm evidence in
the CERN SPS energy range. Tantalising hints were found for medium-size nuclei in data from
the NA49 experiment which strongly motivate the ongoing scan of the phase diagram by the 
NA61 experiment (see Fig.~\ref{na61_na49_scan}). A search for the CP is also in progress
at the Brookhaven RHIC within the beam energy scan (BES) program.



\section*{Acknowledgments}
  I thank the organisers of ISMD2013 for giving me the opportunity to present
  this report and I am grateful to all members of the NA49 and NA61 collaborations
  for their hard work to obtain these experimental results.


\end{document}